\documentstyle[12pt]{article}

\begin{document}
\title{Acoustic Attenuation in High-$T_c$ Superconductors}
\author{T.\ Wolenski\protect\cite{Hamburg}
and J.\ C.\ Swihart\\ 
Department of Physics\\Indiana University\\Bloomington, IN 47405}
\date{Accepted for Publication, Physica C\\IUCM 95-022}
\maketitle

\begin{abstract}
We analyze the acoustic attenuation rate in high-$T_c$ superconductors,
and find that this method offers an additional way to
examine the anisotropy of the superconducting order pa\-ra\-me\-ter in these
materials. We argue that it should be possible to distinguish 
the electronic contribution to the acoustic attenuation, which has a
strong temperature dependence near $T_c$, from the lattice contribution, 
which does
not show a strong temperature dependence near $T_c$. We propose that
this can be utilized to measure the anisotropy of the order parameter
by measuring the attenuation rate near $T_c$ in different directions.
\end{abstract}

Keywords: anisotropic superconductor, energy gap, ultrasound attenuation

\section{Introduction}
\label{sec:intro}

The symmetry of the order parameter $\Delta_{\bf q}$---the energy
gap in the quasiparticle excitation spec\-trum---in the high-$T_c$
superconductors (HTSC's) has been a very active area of research
in the last few years. Assuming that the order parameter is a
spin singlet as seems to be indicated by Knight-shift 
experiments \cite{Barrett}, the angular pairing state has to be
even, leaving $s$-wave ($L$=0) and $d$-wave ($L$=2) as the main
alternatives. Different theories make different predictions for the order
parameter, which is the reason for the interest in the symmetry
of $\Delta_{\bf q}$.

There are essentially two classes of experiments to determine the
symmetry of the energy gap, those that are sensitive to the 
magnitude of the order
parameter, and those that are sensitive to the phase of it.
Measurements of the acoustic
attenuation \cite{MorseBohm} and the NMR nuclear-spin relaxation 
rate \cite{Hebel}, which belong to the first category, were instrumental
in establishing the BCS theory in the conventional 
superconductors \cite{Schrieffer}. Acoustic Attenuation was
also used to examine anisotropy in the conventional superconductors
due to the crystal symmetry \cite{Morse}.

However, while many NMR experiments have been performed on the HTSC's
(cf. the review by Slichter {\em et al.}\ \cite{Slichter93}), acoustic
attenuation experiments have been neglected in recent years. 
We will discuss the
theory of acoustic attenuation for the HTSC's, and propose an
experiment to probe the symmetry of the order parameter using acoustic
attenuation.

Acoustic attenuation is observed with an experimental setup \cite{Rayne}, 
where an ultrasonic signal, typically with frequencies ranging from 
100 MHz to 10 GHz, is fed into a sample through a 
transducer quartz. The resulting phonons with wave vector $\bf q$,
energy $\omega$, and polarization $\lambda$ can be scattered 
by quasiparticles, and the remaining
fraction of phonons is observed at the other end of the sample.
In addition to the electronic mechanism, where the phonons are scattered
by quasiparticles, there will also be a lattice contribution to the
attenuation rate. 
We will discuss how the electronic contribution can be resolved against
the lattice background.

\section{Acoustic Attenuation}
\label{sec:acat}

We consider processes in which phonons are 
absorbed by a sample through scattering of quasiparticles from 
a state ${\bf p_1}$ into a state ${\bf p_2}$, 
where ${\bf q} = {\bf p_2} - {\bf p_1} + {\bf K}$ is 
the momentum of the incoming phonon and ${\bf K}$ is a reciprocal lattice 
vector. The inverse process of spontaneous 
emission of phonons by quasiparticles also has to be taken into account. 

To understand the effect of anisotropy of the order parameter 
it is important to understand how the involved momenta are related. 
The quasiparticle is scattered from a state $\bf p_1$ near the 
Fermi surface to another state $\bf p_2$, 
which also has to be near the Fermi surface. 
This means that both ${\bf p_1}$ and ${\bf p_2}$ are of order $k_F$. 
$\bf q$, the phonon momentum, is smaller by some orders of magnitude. 
This means that $\bf p_1$ and $\bf p_2$ point essentially in the 
same direction. 

The HTSC's are quasi-two-dimensional layered materials. 
Their Fermi surface is nearly cylindrical with little dependence 
on the coordinate in the $c$-direction. 
For a three-dimensional Fermi sphere, $\bf p_1$ and $\bf p_2$ 
are fixed on a belt around the Fermi surface \cite{Morse}, 
perpendicular to the direction of the phonon momentum $\bf q$. 
For a given phonon momentum $\bf q$ in the $a$-$b$-plane in the 
quasi-two-dimensional case, this belt degenerates to two 
points on opposite sides of the Fermi surface.
Thus,  
the order parameter is probed only in a specified direction. 
This provides an opportunity to measure the anisotropy in the 
magnitude of the order parameter. 


The interaction of phonons with quasiparticles can be written as
\begin{equation}
{\cal H}_{\mbox{el-ph}} = \sum_{{\bf p_1},{\bf p_2},s,\lambda} 
g_{{\bf p_1,p_2},\lambda} 
(a_{{\bf q} \lambda} + a_{-{\bf q} \lambda}^\dagger) 
c_{{\bf p_2} s}^\dagger 
c_{{\bf p_1} s},
\label{eq:HamiltonianAc}
\end {equation}
where the relation between $\bf q$, $\bf p_1$, and $\bf p_2$ was given
above. Here $g_{{\bf p_1,p_2},\lambda}$ is the 
interaction strength, $a_{{\bf q} \lambda}$ is a phonon destruction operator,
and $c_{{\bf p_1} s}$ an electron destruction operator. The electron operators 
$c$, $c^\dagger$ can be transformed into superconducting 
quasiparticle operators $\gamma$, 
$\gamma^\dagger$ with a standard Boguljubov transformation.

Starting from this Hamiltonian it is straightforward to derive an
expression for the acoustic attenuation rate (e.g.\ cf.\ 
Schrieffer \cite{Schrieffer}) 
$\alpha_{{\bf q},\lambda}$ with
\begin {equation}
\alpha_{{\bf q},\lambda}=4\pi \sum_{{\bf p_1},{\bf p_2}} |g_{{\bf 
p_1,p_2},\lambda}|^2 \, n^2 ({\bf p_1},{\bf p_2}) \, (f_{\bf p_1}-f_{\bf p_2}) 
\, \delta(E_{\bf p_2}-E_{\bf p_1}-\omega_{{\bf q} \lambda})
\end{equation}
for phonons of wave vector $\bf q$ and polarization $\lambda$. The involved
phonon frequency $\omega_{{\bf q} \lambda}$ is given by the dispersion 
relation of the phonons. 
The square of the so-called coherence factor $n$ is evaluated to be 
\begin {equation}
n^2({\bf p_1},{\bf p_2})=(u_{\bf p_1} u_{\bf p_2} - v_{\bf p_1} v_{\bf 
p_2})^2=
\frac {1} {2} \left(1+ \frac {\epsilon_{\bf p_1} \epsilon_{\bf 
p_2}-\Delta_{\bf p_1} \Delta_{\bf p_2}} {E_{\bf p_1} E_{\bf p_2}}\right).
\end{equation}
Here, ${\epsilon_{\bf p}}$ is the single-particle energy relative 
to the Fermi energy and $E_{\bf p}$ is the superconducting 
quasiparticle excitation energy, 
$E_{\bf p}=\sqrt{\epsilon_{\bf p}^2+ \Delta_{\bf p}^2}$.
$f_{\bf q} \equiv 1/(1+\exp (E_{\bf q}/k_B T))$ is the Fermi function.
We assume that $g_{{\bf p_1,p_2} ,\lambda}$ only depends on the momentum 
transfer ${\bf q} = {\bf p_2} - {\bf p_1}$.

It is assumed that the order parameter depends only on the direction, 
but not on the magnitude of {\bf p}, and that the angular dependence of
the order parameter does not change with temperature $T$.

Finally we are led to 
\begin {eqnarray}
\alpha_{{\bf q},\lambda} & = & \mbox{const.} \times \int d\epsilon_{\bf p_1} 
d\epsilon_{\bf p_2} \left(1-\frac {\Delta_{\bf p_1} \Delta_{\bf p_2}} 
{E_{\bf p_1} 
E_{\bf p_2}}\right) \nonumber \\*
& & \times (f(E_{\bf p_1})-f(E_{\bf p_2})) 
\delta(E_{\bf p_2}-E_{\bf p_1}-\omega_{q \lambda}) \nonumber \\*
& = & \mbox{const.} \times \int dE_{\bf p_1} \frac {E_{\bf p_1}} 
{\sqrt{E_{\bf p_1}^2 
-\Delta_{\bf p_1}^2}} \frac {E_{\bf p_1}+\omega_{q \lambda}} 
{\sqrt{(E_{\bf p_1}+\omega_{q \lambda})^2-\Delta_{\bf p_2}^2}}  
\nonumber 
\\*
& & \times \left(1-\frac {\Delta_{\bf p_1} \Delta_{\bf p_2}} {E_{\bf p_1} 
(E_{\bf p_1}+\omega_{q \lambda})}\right) 
(f(E_{\bf p_1})-f(E_{\bf p_1}+\omega_{q 
\lambda})).
\label{eq:int}
\end{eqnarray}
We do not get an angular integration because $\bf q$ picks certain
values for the directions of $\bf p_1$ and $\bf p_2$.
These directions, $\theta_1$ and $\theta_2$, are very close to one another
due to momentum conservation as argued above, and can be controlled
experimentally.  For our calculations we used 
$\hbar \omega_{q \lambda} = 10^{-5}k_B T_c$.  The results are quite
insensitive to the value of $\omega$ as long as 
$\hbar \omega << k_B T_c$.




\section{Symmetry of the Energy Gap}
\label{sec:symmetry}

To calculate the acoustic attenuation rate $\alpha_{{\bf q},\lambda}(T)$, 
the temperature dependence 
of the reduced gap, $\Delta_0(T)/\Delta_0(0)$, will be assumed to be BCS-like 
as a function of reduced temperature.
We have tested that this assumption is consistent with both an $s$- and 
$d$-wave gap with appropriate potentials in the BCS gap 
equation \cite{Diplom}.  
For the angular dependence of the gap, $\Delta({\theta})$, different 
models will be examined:
\begin {equation}
\Delta(\theta,T) =
\left\{ \begin{array} {ll}
        \Delta_0(T) &\mbox {isotropic } s\mbox{-wave},\\
        \Delta_0(T) \cos (2\theta) &d\mbox{-wave},\\
        \Delta_0(T) \left[ a\cos^2 (2\theta)+(1-a) \right] &\mbox 
{anisotropic            } s\mbox{-wave}.
\end{array} \right.
\label{eq:gapforms}     
\end {equation}


In conventional superconductors one calculates the
attenuation rate in the superconducting state normalized to 
the attenuation rate in the normal state. In HTSC's, however,
this is not an interesting quantity because it is not experimentally
accessible. In conventional superconductors one can always drive
the system into the normal state even at temperatures much below $T_c$
by applying a sufficiently large magnetic field. In HTSC's, however,
the critical fields are prohibitively high. Thus, we normalize the
attenuation rate to its value at $T_c$.


Evaluating Eq.\ (\ref{eq:int}) with $\bf q$ pointing in different 
directions relative to the lattice amounts to taking different
effective magnitudes of $\Delta$. 
We consider $q/k_F$ very small so that
$\theta_1\simeq\theta_2$, and thus 
$\Delta_{\bf p_1}\simeq\Delta_{\bf p_2}$. The direction of $\bf q$
can be controlled experimentally. Fig.\ 1 shows the 
attenuation rate as a function of temperature for different effective
magnitudes of the gap. A $2 \Delta_0(T=0)/k_B T_c=3.5$ corresponds to the
isotropic BCS case.
Because for a particular $\bf q$ only the magnitude of the gap in the
direction perpendicular to $\bf q$ enters, Fig.\ 1 holds for
a particular $\bf q$ no matter what the symmetry of the gap is.
This is true as long as the temperature dependence of the gap in that 
direction behaves like the BCS temperature dependence.

At the high temperatures near $T_c$, $T_c \simeq 90$ K, 
lattice contributions to the attenuation rate become important, 
but the strong temperature dependence of 
the electronic contribution near $T_c$ provides an opportunity
to still measure it.
Since the superconducting
transition should not affect any but the electronic contribution
to the attenuation, measuring the attenuation rate just 
above and below $T_c$ allows one to separate 
{\em electronic} contributions from {\em lattice} contributions.
In particular any eventual anisotropy of a lattice contribution should
not be influenced by the superconducting transition, at least as long as
the temperature difference between the measurements is not too large.
Using this approach, the anisotropy of the electronic attenuation rate
becomes an experimentally accessible quantity, which can be used to
examine the symmetry of the superconducting gap in the HTSC's.

For an isotropic $s$-wave gap the electronic contribution to the 
attenuation rate should not change as the 
crystal is rotated. For an anisotropic order parameter, either 
$s$- or $d$-wave, maxima should be observed 
in directions perpendicular to
where the order parameter 
is a minimum or even has nodes. At a node the attenuation rate should 
go up to the normal state value at the corresponding temperature.

Fig.\ 2 shows how the acoustic attenuation rate varies
at a temperature of 0.95 $T_c$ for different symmetries of the
order parameter as a function of the direction of the incoming phonons
relative to the lattice. A value of $2\Delta_0(T=0)/k_B T_c=6$ is assumed.
All rates are normalized to the acoustic
attenuation rate without a gap, which is essentially the electronic 
attenuation rate at $T_c$, since normal electronic contributions should not 
vary much over
this small range of temperatures. While all symmetries show a significant 
suppression in certain directions, which allow the resolution of 
the electronic contribution against the lattice background, the very
anisotropic symmetries show little or no suppression in those 
directions where minima or nodes of the gap are located.



\section{Discussion and Summary}
\label{sec:disc}

The acoustic attenuation method, which was very successful in 
verifying BCS theory for conventional superconductors, has the 
potential to provide useful information on the order parameter in 
high $T_c$ materials. 

Early on some measurements have been made 
(Yusheng {\em et al.}\ \cite {Yusheng}), 
but it has been argued \cite {Almond} that the effect seen was too large
to be the actual electronic contribution to the attenuation rate.
However, since then sample qualities have been improved considerably,
and, with our proposed focusing on temperatures near
$T_c$, this method has a potential that has not been exploited yet. 

Won and Maki \cite {Won} recently also discussed acoustic attenuation
in HTSC's;
however, they do make some additional approximations to solve the 
problem analytically, only consider $d$-wave, and argue that at {\em low}
temperatures, where the rate is already strongly suppressed, 
this rate should be strongly anisotropic. 

We propose that the attenuation be measured at temperatures 
near $T_c$, where one should see the anisotropy, but still
have a measurable rate. Any observed sharp drop
in the attenuation rate near $T_c$ can be attributed to the 
electronic properties of the system since the lattice properties 
should not change dramatically near the superconducting transition.  

{\bf Note Added in Proof:} After the present paper was submitted for 
publication, we learned of a paper by Kostur {\it et al.} \cite{Kostur}   
(KBF) which reports on calculations of
ultrasonic attenuation in a model d-wave superconductor.
The results are quite similar to those presented here.  We feel that our 
paper compliments KBF in that we consider anisotropic s-wave
superconductors in addition to d-wave.  Also, we suggest that the best 
way to separate the
anisotropic attenuation due to quasiparticles from that due to the
lattice is to compare results just above $T_c$ with those just
below $T_c$, whereas KBF seem to advocate looking at low temperatures.

\section*{Acknowledgements}

We would like to thank J.\ Appel and C.\ Timm for useful discussions. 
T.\ W.\ likes to thank Indiana University for the hospitality
during his stay in Bloomington, and gratefully acknowledges financial support 
through the Indiana University Overseas Exchange Fellowship.

\pagebreak

\pagebreak

\noindent
Figure 1: Acoustic attenuation rate for an effective 
order parameter for different values of $2\Delta_0/kT_c$. Here the 
zero-temperature $\Delta_0$ corresponds to a gap in a particular 
direction.

\bigskip
\bigskip
\noindent
Figure 2: Angular dependence of the attenuation rate for different 
symmetries at a temperature $T=0.95 \, T_c$ with $2\Delta_0(T=0)/k_B T_c=6$,
where $\Delta_0(T)$ is from one of the expressions in Eq.\ (5).
All rates are normalized to the attenuation rate without a gap.

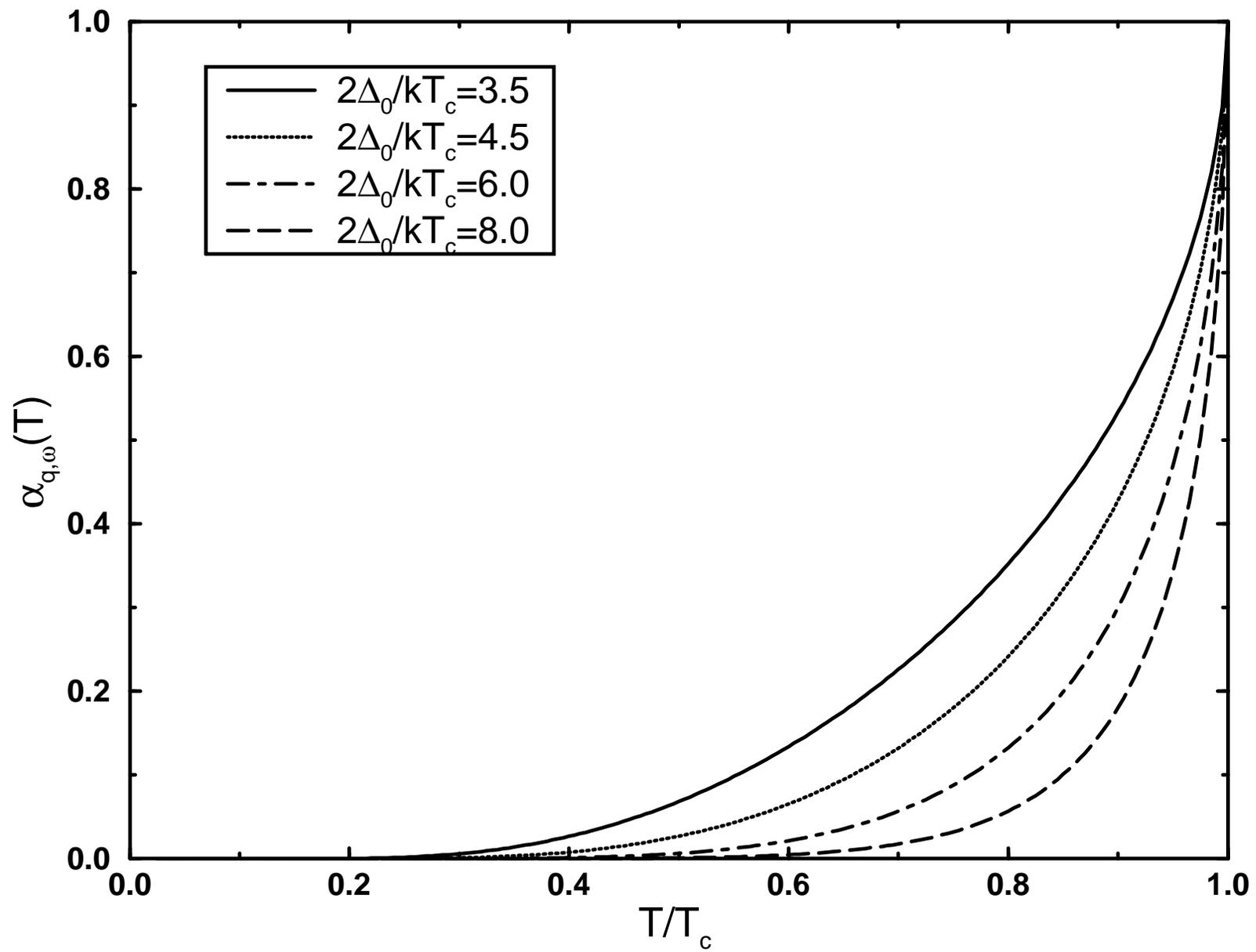
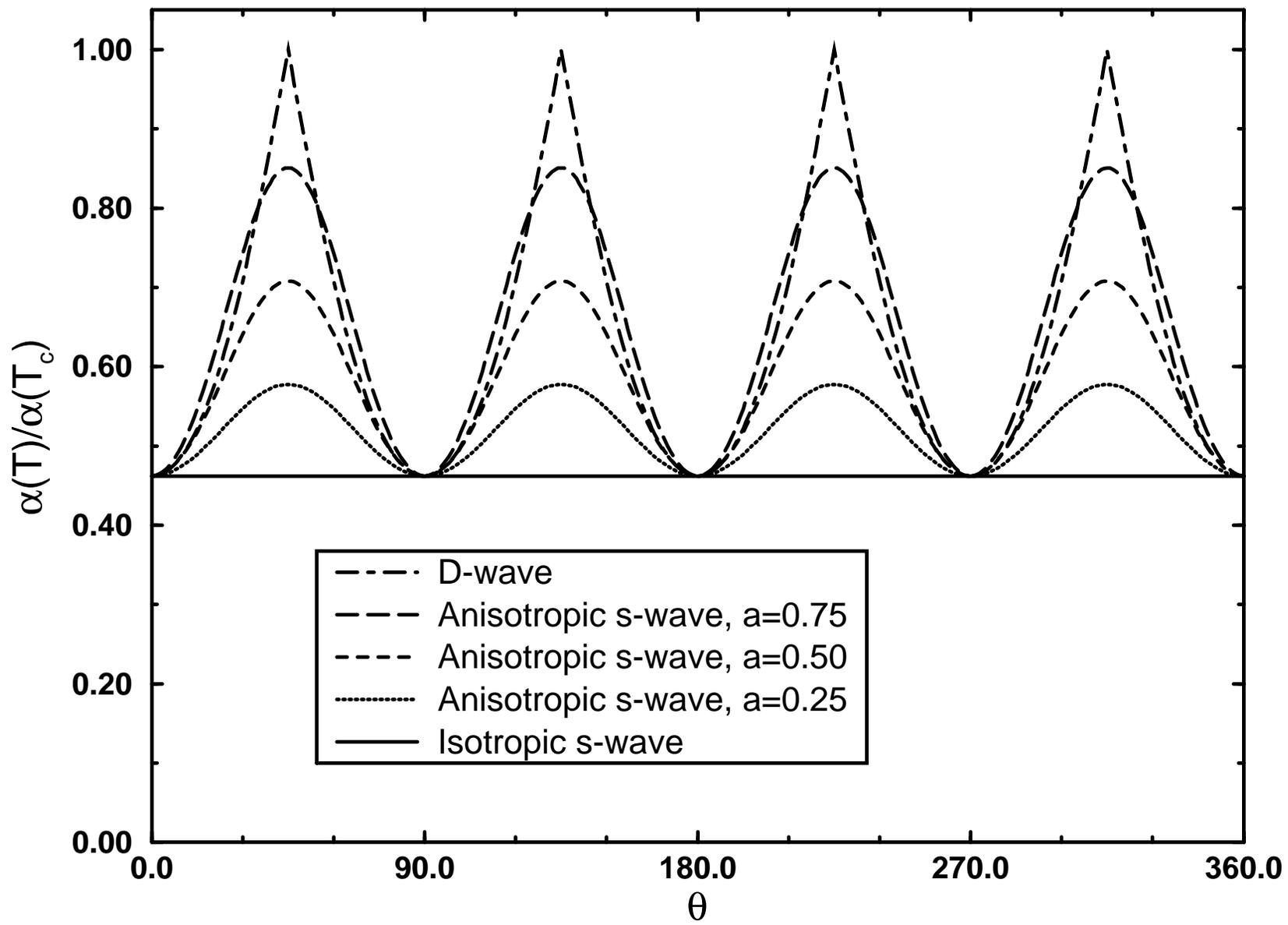
\end{document}